\documentclass[11pt,a4paper]{article}
\pdfoutput=1

\usepackage{jheppub}
\usepackage{physics}
\usepackage{tensor}
\usepackage{amsmath,amsfonts,amssymb}
\usepackage{url}
\usepackage{comment}

\numberwithin{equation}{section}

\newcommand{\fft}[2]{\frac{#1}{#2}}
\newcommand{\ft}[2]{{\textstyle\frac{#1}{#2}}}
\newcommand{\nn}{\nonumber}

\preprint{LCTP-22-14}

\title{T-duality building blocks 
for $\alpha'$ string corrections}

\author[a]{Marina David}
\emailAdd{marina.david@kuleuven.be}

\author[b]{James T. Liu}
\emailAdd{jimliu@umich.edu}

\affiliation[a]{Instituut voor Theoretische Fysica, KU Leuven Celestijnenlaan 200D, B-3001 Leuven, Belgium}

\affiliation[b]{Leinweber Center for Theoretical Physics, University of Michigan, Ann Arbor, MI 48109, U.S.A.}

\abstract{T-duality has been shown to constrain the higher derivative corrections of string theory. We revisit the problem of understanding the T-duality constraints imposed on the $\alpha'$ corrections using the language of a torsionful connection. We find a convenient way to express the $O(d,d)$ invariants in terms of linear combinations of the metric and B-field for general $d$-dimensional torus compactifications, which we then use to revisit the heterotic and bosonic ten-dimensional string action at order $\alpha'$. We also comment on the four-point functions for corrections of order $\alpha'^3$ in the type II string.}


\arxivnumber{}

\begin{document}

\maketitle

\section{Introduction}

An outstanding problem in perturbative string theory is to understand the structure of the higher derivative terms that appear as a series expansion in the string coupling $\alpha'$. Since technical challenges arise at higher orders in $\alpha'$, it becomes crucial to investigate if there are any additional hidden structures that make the derivative corrections more manifest. This would, for example, be useful for the $\alpha'^{3}$ corrections in the type II string, as the full structure of these eight derivative couplings is still unknown.

One common approach is to take advantage of the manifest symmetries to constrain the higher derivative terms. This includes making use of supersymmtry \cite{Green:1998by,Cederwall:2000ye,Peeters:2000qj,Gates:2001hf,Nishino:2001mb,Cederwall:2004cg,Rajaraman:2005ag,Paulos:2008tn,Becker:2017zwe,Becker:2021oiz}, S-duality \cite{Green:1997tv,Green:1997di,Green:2019rhz}, and, what is relevant to this work, T-duality. T-duality appears when compactifying on a circle, or more generally some $d$-dimensional torus. This then constrains the particular background to be $O(d,d)$ invariant. Restricting to the NS-NS sector, the tree level couplings are indistinguishable between type IIA and type IIB string theories and T-duality invariance is universal at tree level in the type II string. At $\mathcal{O}(\alpha')$ in the bosonic and heterotic string, the T-duality transformation rules were explored in \cite{Bergshoeff:1995cg,Kaloper:1997ux} and T-duality invariants were constructed in \cite{Marques:2015vua,Baron:2017dvb} using Double Field Theory (DFT). Moreover, T-duality can be used to compute the corrections to black hole solutions as was done in \cite{Cano:2018qev,Edelstein:2018ewc,Cano:2018brq,Edelstein:2019wzg,Elgood:2020xwu,Ortin:2020xdm,Cano:2021nzo,Cano:2022tmn}. For the type II string, the higher derivative corrections at order $\mathcal{O}(\alpha'^{3})$ have been investigated in \cite{Garousi:2012yr,Garousi:2012jp,Garousi:2013zca,Liu:2013dna,Liu:2019ses} and completed via T-duality in  \cite{Garousi:2020gio,Garousi:2020lof}.

An additional computational advantage arises when performing a `cosmological reduction' in which all but the temporal dimension are compactified \cite{Meissner:1996sa,Godazgar:2013bja,Hohm:2015doa}. The only fields that survive in this reduction are scalars, so this gives us an opportunity to study the scalar sector where we have a nonzero dilaton, and internal metric and B-field components $g_{ij}$ and $b_{ij}$. This, for example has been investigated in \cite{Codina:2020kvj}, where it was then deduced what the eight-derivative couplings quartic in the Riemann tensor must be. This framework was then extended in the context of a torsionful connection \cite{David:2021jqn}, where the $O(1,1)$ invariants can be written as linear combinations of the metric and B-field defined by a matrix $N_{\pm}=g^{-1}\partial_t(g\pm b)$. Then, the T-duality invariant quantities are traces of even powers of products of the matrices $N_{\pm}$ with alternating signs. We may then wonder if this structure appears for a $d$-dimensional compactification, where there is a natural generalization of $N_\pm$ to $N_{\mu\,\pm}\equiv g^{-1}\partial_\mu(g\pm b)$.

We revisit the higher derivative corrections at order $\mathcal{O}(\alpha')$ and $\mathcal{O}(\alpha'^{3})$ to investigate the hidden structures of string theory for $d$-dimensional compacification. This is similar to the work in \cite{Eloy:2020dko}, although here we use the language of a torsionful connection. In this setup, we can formulate building blocks that appear as $O(d,d)$ invariant quantities allowing us to study the additional structures of the string action that may become manifest. We find as expected that we can characterize the T-duality invariants formed out of first derivatives of $g_{ij}$ and $b_{ij}$ as products of the generalized $N_{\mu\,\pm}$ with alternating signs. However, the situation for invariants constructed out of higher derivatives of $g_{ij}$ and $b_{ij}$ is less clear since the `alternating signs rule' may no longer apply as some of the traces contain a product of odd numbers of $N_{\mu\,\pm}$ matrices.

With the building blocks we construct, we revisit the first order $\alpha'$ corrections for the heterotic and bosonic string action. We find that indeed we can write the entire first order corrections as traces of products of the $N_{\mu\,\pm}$ matrices with alternating signs, generalizing the results in the case of the cosmological reduction \cite{David:2021jqn}. However, one main difference appears in that the linear combination of the T-duality building blocks is not equivalent for the heterotic and bosonic string, although this distinction vanishes for the reduction to one temporal dimension.  This provides an example where the $d$-dimensional reduction provides more information on the T-duality invariants than a cosmological reduction alone would.

There is currently limited literature on string amplitudes for various $n$-point functions for $n\ge5$, mainly due to technical challenges. However, even in the cosmological reduction, we are still able to extract valuable information about the couplings which can then be matched via higher-point string amplitudes. For example, it has been shown in \cite{David:2021jqn} that T-duality in the cosmological reduction can constrain the five-point function of the form $H^2R^3$, which were obtained via tree-level string amplitudes \cite{Liu:2019ses}. In this example, there are eight coupling constants undetermined by the four-point function, and five of them can be explicitly fixed via T-duality, leaving three unconstrained \cite{David:2021jqn}.  More generally, using DFT inspired objects, both $H^2R^3$ and $H^2\nabla H^2 R$ were fully constrained by $O(d,d)$ invariance in \cite{Wulff:2021fhr}.  We may also wonder to what extent can we learn about the four and five point functions with the newly constructed building blocks in the context of the $d$-dimensional reduction.

This paper is organized as follows. In Section~\ref{section: the torus reduction}, we review the torus reduction using the framework of a torsionful connection, and introduce notation useful for constructing T-duality building blocks.  In Section~\ref{section: alpha prime corrections}, we make use of these building blocks to compute the $\mathcal O(\alpha')$ corrections to the heterotic and bosonic strings. We then turn to the $\mathcal O(\alpha'^3)$ corrections to the type II string and make comments on the structure of the four and five point couplings in Section~\ref{section: alpha prime cubed corrections} and end with concluding remarks and open directions in Section~\ref{section: discussion}.

\section{The torus reduction with a torsionful connection} \label{section: the torus reduction}

In this section, we review the $d$-dimensional torus reduction in the context of a torsionful connection. This is a direct generalization of the cosmological reduction with nonzero torsion \cite{David:2021jqn} and therefore, as such we find a convenient way to express the $O(d,d)$ invariants in terms of linear combinations of the metric and B-field. We then discuss several identities that become useful in constructing the higher derivative corrections.

\subsection{The torus reduction}

We begin with a review of the reduction of the NSNS sector of string theory on a $d$-dimensional torus.  The fields we consider are the metric $g_{MN}$, B-field $B_{MN}$ and ten-dimensional dilaton $\phi$, with string frame two-derivative NSNS sector Lagrangian
\begin{equation}
    E^{-1}\mathcal L_{10}=e^{-2\phi}\left[R+4\partial_M\phi\partial^M\phi-\fft1{12}H_{MNP}H^{MNP}\right].
\end{equation}
We take a standard torus reduction
\begin{align}
    ds^2&=g_{\mu\nu}dx^\mu dx^\nu+g_{ij}Dy^iDy^j,\qquad Dy^i=dy^i+A_\mu^idx^\mu,\nn\\
    B&=\ft12b_{\mu\nu}dx^{\mu} dx^\nu+B_{\mu i}dx^\mu Dy^i+\ft12b_{ij}Dy^i\wedge Dy^j,\nn\\
    \phi&=\Phi+\ft14\log\det g_{ij},
\label{eq:treduct}
\end{align}
where the Latin indices $(i,j, \dots)$ denote the $T^{d}$ indices and the Greek indices $(\mu,\nu, \dots)$ are the indices of the lower-dimensional spacetime.  Here $\Phi$ is the lower-dimensional dilaton.  We use capital Latin indices $(M,N,\dots)$ to denote the $D$-dimensional indices.  This metric admits a natural vielbein basis
\begin{equation}
    ds^2=\eta_{\alpha\beta}E^\alpha E^\beta+\delta_{ab}E^aE^b,
\end{equation}
where
\begin{equation}
    E^\alpha=e_\mu^\alpha dx^\mu,\qquad E^a=e_i^a(dy^i+A_\mu^idx^\mu).
\end{equation}
While the vector fields are an integral part of the reduction, we focus only on T-duality in the \textit{scalar sector}, and hence set $A_\mu^i=0$ from now on.  In this case, the two-derivative Lagrangian reduces to
\begin{equation}
    e^{-1}\mathcal L_{10-d}=e^{-\Phi}\left[R+\partial_\mu\Phi\partial^\mu\Phi-\fft12h_{\mu\nu\rho}h^{\mu\nu\rho}+\fft18\Tr(\partial_\mu\mathcal H\eta\partial^\mu\mathcal H\eta)\right],
\label{eq:redlag}
\end{equation}
where $\mathcal{H}$ is the $(2d) \times (2d)$ scalar matrix
\begin{equation}
    \mathcal H=\begin{pmatrix}g^{-1}&-g^{-1}b\\bg^{-1}&g-bg^{-1}b\end{pmatrix},
\label{eq:calH}
\end{equation}
and $\eta$ is the $O(d,d)$ metric
\begin{equation}
   \eta =\begin{pmatrix}0 & \mathbf{1} \\ \mathbf{1} & 0 \end{pmatrix}.
\end{equation}
T-duality invariance of this two-derivative Lagrangian is well established.

Higher-derivative corrections to the two-derivative Lagrangian are built out of gauge-invariant combinations of the Riemann tensor, three-form field strength $H$ and derivatives of the dilaton $\phi$.  The torus reduction of $H$ and $\phi$ is straightforward, leaving only Riemann to consider.  For this, we choose to use a torsionful connection given by
\begin{equation}
    \Omega_\pm=\omega\pm\ft12 H=\omega^{AB}\pm\ft12H_M{}^{AB}dx^M,
\end{equation}
where $H$ is the 3-form field strength, $H=dB$. The torus reduction of $\Omega_{\pm}$ is straightforward and, with $A_\mu^i=0$, we find
\begin{equation}
    \Omega_\pm^{AB}=\begin{pmatrix}\omega_\pm^{\alpha\beta}&-\fft12e^{ib}\partial^\alpha(g_{ij}\mp b_{ij})dy^j\\
    \fft12e^{ia}\partial^\beta(g_{ij}\mp b_{ij})dy^j&\fft12(e^i_ade_{ib}-e^i_bde_{ia}\pm e^{ia}e^{jb}db_{ij})\end{pmatrix}.
\label{eq:Omegapm}
\end{equation}
We can then compute the curvature of this torsionful connection given by
\begin{equation}
    R_{MN}{ }^{PQ}\left(\Omega_{\pm}\right)=R_{MN}{ }^{PQ}(\Omega) \pm \nabla_{[M} H_{N]}{}^{PQ}+\frac{1}{2} H_{[M}{}^{PR} H_{N]R}{}^{Q}.
\end{equation}
The components of the Riemann tensor are then
\begin{align}
    R_{\rho\sigma}{}^{\mu\nu}(\Omega_\pm)&=R_{\rho\sigma}{}^{\mu\nu}(\omega_\pm),\nn\\
    R_{ij}{}^{\mu\nu}(\Omega_\pm)&=-\ft14(g_{ik}N_\pm^{\mu k}{}_lN_\mp^{\nu l}{}_j-g_{jk}N_\pm^{\mu k}{}_lN_\mp^{\nu l}{}_i),\nn\\
    R_{\mu\nu}{}^{ij}(\Omega_\pm)&=-\ft14(N_{\mu\mp}^i{}_kN_{\nu\pm}^k{}_l-N_{\nu\mp}^i{}_kN_{\mu\pm}^k{}_l)g^{lj},\nn\\
    R_{\nu j}{}^{\mu i}(\Omega_\pm)&=-\ft14(2\nabla^{(\pm)}_\nu N_\mp^{\mu i}{}_j+N_{\nu\pm}^i{}_kN_\mp^{\mu k}{}_j),\nn\\
    R_{kl}{}^{ij}(\Omega_\pm)&=-\ft14(N_\mp^{\mu i}{}_kN_{\mp\mu}^j{}_l-N_\mp^{\mu i}{}_lN_{\mp\mu}^j{}_k),
\label{eq:Riemann}
\end{align}
where we have defined the matrix
\begin{equation}
    N_{\mu\pm}^i{}_j=g^{il}\partial_\mu(g_{lj}\pm b_{lj}).
\end{equation}
Note that the covariant derivative $\nabla_\nu^{(\pm)}$ is with respect to the torsionful connection.  In particular
\begin{equation}
    \nabla^{(\pm)}_\nu N_\mp^{\mu i}{}_j= \nabla_\nu N_\mp^{\mu i}{}_j\pm\ft12h_\nu{}^\mu{}_\rho N_\mp^{\rho i}{}_j.
\label{eq:nabla+}
\end{equation}

We recall that the symmetries of $R_{AB}{}^{CD}(\Omega_{\pm})$ differ from its torsionfree counterpart. More specifically, the first pair and the second pair are both separately antisymmetric, while pairwise interchanges gives
\begin{align}
    R_{AB}{}^{CD}(\Omega_{\pm}) &= R^{CD}{}_{AB}(\Omega_{\mp}).
\end{align}
One of our main observations is that the torsionful Riemann tensor and T-dual invariant quantities can be written in terms of $N^i_{\mu\pm j}$. Note that this generalizes the $N_\pm^i{}_j=g^{il}(\dot g_{lj}\pm\dot b_{lj})$ of the cosmological reduction \cite{David:2021jqn}, in which the $O(d,d)$ invariants were composed of traces of products of $N$'s with alternating signs.

\subsection{Forming T-duality invariants}

As demonstrated above, torus reduced higher-derivative couplings can be written in terms of $N^i_{\mu\pm j}$, $\Phi$ and $h=db$ as well as the lower-dimensional Riemann tensor and the field strengths $F^i=dA^i$.  Focusing on the $O(d,d)$ scalars, we only concern ourselves with invariants built out of $N^i_{\mu\pm j}$.  However, this $N$ matrix by itself does not have any obvious $O(d,d)$ transformation properties.

To make contact with the $O(d,d)$ invariants, we define the $O(d,d)$ matrix
\begin{equation}
    \mathcal S=\eta \mathcal H=\begin{pmatrix}bg^{-1}\quad&g-bg^{-1}b\\g^{-1}&-g^{-1}b\end{pmatrix},
\end{equation}
where $\mathcal H$ is given in (\ref{eq:calH}).  This matrix transforms as
\begin{equation}
    \mathcal S\to \mathfrak g^{-1}\mathcal S\mathfrak g,\qquad\mbox{where}\qquad\mathfrak g\eta\mathfrak g^t=\eta,
\end{equation}
under $O(d,d)$ transformations.  Hence invariants are formed by taking traces of products of $\mathcal S$ and its derivatives. Note, however, that $\mathcal S$ has the property that
\begin{align} \label{eq: S squared equals 1}
    \mathcal S^2=1,
\end{align}
which gives us insight on the types of invariants we can build out of $\mathcal{S}$ and its derivatives. First, there can be at most one $\mathcal S$ separating derivative terms inside the trace. Furthermore, from $\mathcal S^2=1$, we can form the projections
\begin{equation}
    P_\pm=\ft12(1\pm\mathcal S).
\end{equation}
Expanding out $\partial_\mu(\mathcal S^2)=0$ gives $\mathcal S\partial_\mu\mathcal S=-(\partial_\mu\mathcal S)\mathcal S$, which is equivalent to $P_+\partial_\mu\mathcal S=\partial_\mu\mathcal SP_-$.  This shows that $(P_+\partial_\mu\mathcal S)^2=0$, which in turn allows us to transform $P_+\partial_\mu\mathcal S$ into an upper triangular matrix.  Working this out, we find the elegant result
\begin{equation}
    \partial_\mu\mathcal S=W\begin{pmatrix}&N_{\mu-}\\-N_{\mu+}\end{pmatrix}W^{-1},\qquad
    \mathcal S\partial_\mu\mathcal S=W\begin{pmatrix}&N_{\mu-}\\N_{\mu+}\end{pmatrix}W^{-1}.
\end{equation}
where
\begin{equation}
    W=\fft1{\sqrt2}\begin{pmatrix}(g+b)\quad&-(g-b)\\1&1\end{pmatrix}.
\end{equation}

This relation between $\partial_\mu\mathcal S$ and $N_{\mu\pm}$ allows us to completely characterize single trace invariants formed out of $\mathcal S$ and $\partial_\mu\mathcal S$. Starting with single derivatives only, we have
\begin{equation}
    \Tr(\partial_{\mu_1}\mathcal S\partial_{\mu_2}\mathcal S\cdots\partial_{\mu_{2n}}\mathcal S)=(-1)^n[\Tr(N_{\mu_1-}N_{\mu_2+}\cdots N_{\mu_{2n}+})+\Tr(N_{\mu_1+}N_{\mu_2-}\cdots N_{\mu_{2n}-})].
\label{eq:dSone}
\end{equation}
If we replace any number of the $\partial_\mu\mathcal S$ terms with $\mathcal S\partial_\mu\mathcal S$, we flip the sign of the corresponding $N_{\mu+}$ term in the trace.  Depending on how many such sign flips there are, we could flip the sign of the first trace or second trace or both traces in the above.  In the end, there are only two inequivalent possibilities.  So we might as well take only a single $\mathcal S\partial_\mu\mathcal S$ in the trace to get the second invariant
\begin{equation}
    \Tr(\mathcal S\partial_{\mu_1}\mathcal S\partial_{\mu_2}\mathcal S\cdots\partial_{\mu_{2n}}\mathcal S)=(-1)^n[\Tr(N_{\mu_1-}N_{\mu_2+}\cdots N_{\mu_{2n}+})-\Tr(N_{\mu_1+}N_{\mu_2-}\cdots N_{\mu_{2n}-})].
\label{eq:dStwo}
\end{equation}
As an example, for the cosmological reduction, both terms on the right hand side of \eqref{eq:dStwo} are equivalent as all the derivatives become time derivatives. Hence \eqref{eq:dStwo} vanishes in the cosmological reduction case, leaving only the T-duality invariant \eqref{eq:dSone}
\begin{equation} \label{building block 1}
\Tr\left(\dot{\mathcal{S}}^{2 n}\right)=2(-1)^{n} \Tr\left(\left(N_{+} N_{-}\right)^{n}\right), \quad(n \geq 0),
\end{equation}
which was found in \cite{David:2021jqn}.

We now move on to second derivatives on $\mathcal S$, where the invariants explicitly break the alternating signature of the matrices $N$. To see this more clearly, we first summarize the quantities built out of $\mathcal S$ and its derivatives as
\begin{align}
    \mathcal S&=W\begin{pmatrix}1&0\\0&-1\end{pmatrix}W^{-1},\nn\\
    \partial_\mu\mathcal S&=W\begin{pmatrix}0&N_{\mu-}\\-N_{\mu+}&0\end{pmatrix}W^{-1},\nn\\
    \nabla_\mu\nabla_\nu\mathcal S&=W\begin{pmatrix}N_{(\mu-}N_{\nu)+}&\nabla_{(\mu}N_{\nu)-}+Y_{\mu\nu-}\\-\nabla_{(\mu}N_{\nu)+}-Y_{\mu\nu+}&-N_{(\mu+}N_{\nu)-}\end{pmatrix}W^{-1},
\label{eq:Scovars}
\end{align}
where we have defined
\begin{align}
    Y_{\mu\nu \pm} &= \frac{1}{2}(N_{(\mu\mp}N_{\nu)\pm}-N_{(\mu \pm}N_{\nu)\pm}).
\end{align}
We can build invariants by multiplying these quantities together and taking the trace.  However, the difficulty in characterizing these invariants lies in the off-diagonal terms of $\nabla_\mu\nabla_\nu\mathcal S$.  While the diagonal terms follow the alternating $N_{\mu-}N_{\nu+}$ or $N_{\mu+}N_{\nu-}$ pattern of the first derivative invariants, the $Y_{\mu\nu\pm}$ terms involve two $N$'s which breaks the even/odd $N$ structure of the diagonal/off-diagonal entries.  Moreover, the structure of $Y_{\mu\nu\pm}$ explicitly breaks the alternating $\pm$ structure of the $N$'s.  In the absence of a systematic treatment of second derivative of $\mathcal S$ invariants, we consider several examples in Section~\ref{section: alpha prime corrections} when we examine the $\mathcal O(\alpha')$ corrections to the bosonic and heterotic string.


Before moving on, however, we summarize several identities involving second derivatives on $\mathcal{S}$ which translate into $\partial N$ type terms. In order to manipulate these terms, it is useful to consider the transpose relation
\begin{equation}
    g^{-1}(\nabla_\mu N_{\nu\pm})^Tg=\nabla_\mu N_{\nu\mp}+L_\mu N_{\nu\mp}-N_{\nu\mp}L_\mu,
\label{eq:transprel}
\end{equation}
as well as the derivative swapping relation
\begin{equation}
    \nabla_\mu N_{\nu\pm}-\nabla_\nu N_{\mu\pm}=L_\nu N_{\mu\pm}-L_\mu N_{\nu\pm}.
\label{eq:swaprel}
\end{equation}
Here we have defined
\begin{equation}
    L_\mu=g^{-1}\partial_\mu g=\ft12(N_{\mu\,+}+N_{\mu\,-}).
\end{equation}
As for manipulations of $\mathcal S$ and its derivatives, note that
\begin{equation}
    \mathcal S\partial_\mu\mathcal S=-(\partial_\mu\mathcal S)\mathcal S.
\end{equation}
This allows us to move $\mathcal S$ around in single derivative expressions.


\subsection{General approach to verifying T-duality} \label{subsection: General approach to verifying T-duality}

With the above setup in mind, we now turn to the general framework for investigating the T-duality properties of tree-level higher derivative corrections.  The general starting point is to focus on a particular order in the string $\alpha'$ expansion, say $\mathcal O(\alpha')$ in the bosonic or heterotic string or $\mathcal O(\alpha'^3)$ for the type II string.  If the higher derivative couplings are already known, then the reduction serves as a useful consistency check while providing additional geometric insight on the reduction.  On the other hand, if the couplings are undetermined or only partially determined, then one can introduce a complete basis of linearly independent terms modulo field redefinitions.  In this case, T-duality invariance can narrow down the undetermined coefficients and, in some cases, even select out a unique invariant \cite{Garousi:2020gio,Garousi:2020lof}.

In any case, the starting point is a set of higher-derivative couplings in the ten-dimensional Lagrangian.  After reduction on $T^d$, and focusing only on the $O(d,d)$ scalars, the resulting Lagrangian will be written in terms of $N_{\mu\,\pm}$ and its derivatives.  Assuming the ten-dimensional couplings are given by no more than second derivatives, the lower-dimensional Lagrangian then admits an expansion of the schematic form
\begin{equation}
    \mathcal L_{10-d}^{(\partial^{2n})}\sim(\alpha')^{n-1}e^{-\Phi}[(\nabla N)^n+N^2(\nabla N)^{n-1}+N^4(\nabla N)^{n-2}+\cdots+N^{2n}],
\label{eq:NdN}
\end{equation}
where we are considering a $(2n)$-derivative coupling.  In the cosmological reduction case, all non-trivial derivatives are time derivatives, and we may use a combination of equations of motion for $\dot N$ and integration by parts to rewrite the entire expression as $\mathcal L_{1}^{(\partial^{2n})}\sim(\alpha')^{n-1}e^{-\Phi}N^{2n}$, which only involves first derivatives of the scalars \cite{Hohm:2015doa,Codina:2021cxh}.  T-duality invariance then requires that the $N$'s enter as traces of alternating strings of $N_+$ and $N_-$, so as to form invariants of the form \eqref{building block 1} as in \cite{David:2021jqn}.  It is worth emphasizing here that use of the two-derivative equations of motion amounts to a field redefinition, so in fact the basic torus reduction, (\ref{eq:treduct}), necessarily receives higher-order corrections if we wish to retain a manifestly $O(d,d)$ invariant torroidially reduced Lagrangian.

The situation is more complicated for general $d$-dimensional reductions as the derivative $\nabla_\mu N_{\nu\,\pm}$ is no longer an equation of motion, except in the case $\nabla^\mu N_{\mu\,\pm}$.  Thus there does not appear to be a general means of reducing (\ref{eq:NdN}) to eliminate all second-derivatives of the scalars.  This, of course, does not signify a failure of $O(d,d)$, as invariants can be formed out of $\nabla_\mu\nabla_\nu\mathcal S$.  However, the presence of $Y_{\mu\nu\pm}$ in (\ref{eq:Scovars}) complicates the assembly of the $N$ and $\nabla N$ terms into good invariants.

Even without a systematic approach to identifying invariants, it is natural to follow a general approach of working from left to right in the schematic expression (\ref{eq:NdN}).  With a slight abuse of notation, we denote terms of $\mathcal O(N^k)$ as $k$-point functions.  Then the first term in (\ref{eq:NdN}) corresponds to an $n$-point function while the last one corresponds to a $(2n)$-point function.  We start with the $n$-point function and make the identification
\begin{equation}
    (\nabla N)^n\to(\nabla\nabla\mathcal S)^n.
\label{eq:dNtoddS}
\end{equation}
Note that this shifts the higher point functions in (\ref{eq:NdN}) since $\nabla\nabla\mathcal S$ involves both $\nabla N$ and $N^2$.  The reduced Lagrangian now takes the form
\begin{equation}
    \mathcal L_{10-d}^{(\partial^{2n})}\sim(\alpha')^{n-1}e^{-\Phi}[(\nabla\nabla\mathcal S)^n+N^2(\nabla N)^{n-1}+N^4(\nabla N)^{n-2}+\cdots+N^{2n}],
\end{equation}
where the first term is manifestly $O(d,d)$ invariant.  It is worth pointing out that this identification, (\ref{eq:dNtoddS}), requires that $\nabla N_\pm$ enters with alternating signs, which provides an immediate constraint on $O(d,d)$ invariance.

It is now suggestive that we could use a similar approach to rewriting the $(n+1)$-point term as $N^2(\nabla N)^{n-1}\to N^2(\nabla\nabla\mathcal S)^{n-1}\to(\partial\mathcal S)^2(\nabla\nabla\mathcal S)^{n-1}$.  This would then lead to a general algorithm to start with the $n$-point function and rewrite things  order by order until we arrive at the $(2n)$-point function.  At each step along the way, we may find additional constraints imposed by $O(d,d)$ invariance from the requirement of alternating signs on the $N_\pm$.

While this is indeed a reasonable approach in the general sense, there are some important details that complicate the actual implementation of such an algorithm.  One major difficulty is the necessity of integration of parts and use of the equations of motion, as evidenced in the cosmological reduction.  At the $(n+k)$-point function level, the term $N^{2k}(\nabla N)^{n-k}$ could either correspond directly to an $O(d,d)$ invariant $(\partial\mathcal S)^k(\nabla\nabla\mathcal S)^{n-k}$ or could be shifted to a higher $(n+k+1)$-point function of the form $N^{2k+2}(\nabla N)^{n-k-1}$ using integration by parts and the equations of motion, depending on how the spacetime indices are arranged.  Although it may be possible to find a general procedure for splitting $N^{2k}(\nabla N)^{n-k}$ into an $(n+k)$-point invariant plus higher point functions, in practice we have only been able to work with this on a case by case basis.

In the next section, we consider four-derivative invariants for the bosonic and heterotic string.  In this case, we start with
\begin{equation}
    \mathcal L_{10-d}^{(\partial^{4})}\sim\alpha'e^{-\Phi}[(\nabla N)^2+N^2\nabla N+N^4].
\end{equation}
The two-point term $(\nabla N)^2$ is obtained from the reduction of $(R_{MNPQ})^2$, and automatically takes the $O(d,d)$ invariant form
$\Tr(\nabla_\mu N_{\nu\,+}\nabla^\mu N^\nu_-)$, which we rewrite as $(\nabla_\mu\nabla\nu\mathcal S)^2$ along with additional $N^2\nabla N$ and $N^4$ terms.  We then work with the three-point term $N^2\nabla N$ and use field redefinitions to shift it completely to the four-point $N^4$ level.  In particular, we do not introduce any invariants of the form $(\partial\mathcal S)^2\nabla\nabla\mathcal S$.  The resulting expression is then of the form
\begin{equation}
    \mathcal L_{10-d}^{(\partial^{4})}\sim\alpha'e^{-\Phi}[(\nabla_\mu\nabla_\nu\mathcal S)^2+N^4],
\end{equation}
It is now clear from the pattern of the signs of the $N_\pm$ which terms are T-duality invariant and which are not.

Similarly, we can do this at the eight derivative level for the type II string starting at the four-point function level and include higher-point couplings order by order to ensure invariance under $O(d,d)$.  However, without a systematic approach, the number of terms to consider rapidly increases, making it difficult to extract simple T-duality invariant expressions formed out of $\mathcal S$ and its derivatives from the $N$ and $\nabla N$ starting point.

\section{T-duality invariance up to $\mathcal O(\alpha')$} \label{section: alpha prime corrections}

In this section, we consider T-duality invariant expressions built out of traces of $\mathcal S$, $\partial\mathcal S$ and $\nabla\nabla\mathcal S$ up to $\mathcal O(\alpha')$.  In particular, we consider traces of two-derivative terms and four-derivative terms. We then discuss possible field redefinitions and reexamine the heterotic and bosonic string at order $\alpha'$.

\subsection{Two-derivative invariants}

At the two-derivative level, we can generate two elements, $\nabla\nabla\mathcal S$ and $(\partial\mathcal S)^2$, along with any number of undifferentiated $\mathcal S$ insertions.  It is then straightforward to show that
\begin{align}
    \Tr\relax(\nabla_\mu\nabla_\nu\mathcal S)&=0, & \qquad\Tr\relax(\mathcal S\nabla_\mu\nabla_\nu\mathcal S)&=2\Tr(N_{(\mu+}N_{\nu-)}),\nn\\
    \Tr\relax(\partial_\mu\mathcal S\partial_\nu\mathcal S)&=-2\Tr(N_{(\mu+}N_{\nu-)}),& \qquad\Tr\relax(\mathcal S\partial_\mu\mathcal S\partial_\nu\mathcal S)&=2\Tr(N_{[\mu+}N_{\nu-]}).
\end{align}
However, with the transpose property of the trace, we find
\begin{equation}
    \Tr(N_{\mu+}N_{\nu-})=\Tr(N_{\nu+}N_{\mu-})\qquad\Rightarrow\qquad \Tr(N_{[\mu+}N_{\nu-]})=0.
\end{equation}
This indicates that, at the two-derivative level, there is only a single invariant, which we can take to be
\begin{align}
   \Tr\relax(\partial_\mu\mathcal S\partial_\nu\mathcal S)&=-2\Tr(N_{\mu+}N_{\nu-}).
\label{eq:2di}
\end{align}
This invariant corresponds to the last term in the two-derivative Lagrangian in (\ref{eq:redlag}).  Note, also, that this implies the second-derivative invariant, $\Tr\relax(\mathcal S\nabla_\mu\nabla_\nu\mathcal S)$, when expanded, consists only of first-derivatives of the scalars $g_{ij}\pm b_{ij}$.

\subsection{Four-derivative invariants}

At the four-derivative level, we can consider terms of the form $(\nabla\nabla\mathcal S)^2$, $(\nabla\nabla\mathcal S)(\partial\mathcal S)^2$ and $(\partial\mathcal S)^4$ with any number of $\mathcal S$ insertions.  Invariants formed out of $\partial\mathcal S$ only can be read off from (\ref{eq:dSone}) and (\ref{eq:dStwo})
\begin{align}
    \Tr(\partial_\mu\mathcal S\partial_\nu\mathcal S\partial_\rho\mathcal S\partial_\sigma\mathcal S)&=\Tr(N_{\mu-}N_{\nu+}N_{\rho-}N_{\sigma+})+\Tr(N_{\mu+}N_{\nu-}N_{\rho+}N_{\sigma-}),\nn\\
    \Tr(\mathcal S\partial_\mu\mathcal S\partial_\nu\mathcal S\partial_\rho\mathcal S\partial_\sigma\mathcal S)&=\Tr(N_{\mu-}N_{\nu+}N_{\rho-}N_{\sigma+})-\Tr(N_{\mu+}N_{\nu-}N_{\rho+}N_{\sigma-}).
\label{eq:4di1}
\end{align}

For invariants built from $\partial\partial\mathcal S$, we find it convenient to rewrite the last line of (\ref{eq:Scovars}) in the asymmetric form
\begin{align}
    \begin{split} &\nabla_\mu\nabla_\nu\mathcal S\\&=W\begin{pmatrix}N_{(\mu-}N_{\nu)+}&\nabla_{\mu}N_{\nu-}+\fft12(N_{\mu+}N_{\nu-}-N_{\nu-}N_{\mu-})\\-\nabla_{\mu}N_{\nu+}-\fft12(N_{\mu-}N_{\nu+}-N_{\nu+}N_{\mu+})&-N_{(\mu+}N_{\nu)-}\end{pmatrix}W^{-1}.
    \end{split}
\end{align}
As a result, we see that
\begin{equation}
    \Tr(\nabla_\mu\nabla_\nu\mathcal S\nabla_\rho\nabla_\sigma\mathcal S)=-\Tr(\nabla_\mu N_{\nu-}\nabla_\rho N_{\sigma+})-\Tr(\nabla_\mu N_{\nu+}\nabla_\rho N_{\sigma-})+\cdots,
\end{equation}
where we have focused on the $(\nabla N)^2$ terms only.  Taking a transpose of the second term using (\ref{eq:transprel}) then gives
\begin{equation}
    \Tr(\nabla_\mu\nabla_\nu\mathcal S\nabla_\rho\nabla_\sigma\mathcal S)=-2\Tr(\nabla_\mu N_{\nu-}\nabla_\rho N_{\sigma+})+\cdots,
\end{equation}
where $\cdots$ indicates terms of the form $(\nabla N)N^2$ and $N^4$. In fact, what we see is that there are many transformations that can be used to move terms around so many expressions may be redundant or linearly dependent.  It would be useful to choose a canonical basis in some way to eliminate such redundancies.  For example, for $(\nabla\nabla\mathcal S)^2$, we can bring it into the form $\Tr(\nabla N_-\nabla N_+)+\cdots$, while $(\nabla\nabla\mathcal S)(\partial\mathcal S)^2$ can always be brought into the form $\Tr(\nabla N_-N_\pm N_\pm)+\cdots$., where the $\pm$ signs are uncorrelated.

Instead of enumerating all the $(\nabla\nabla\mathcal S)^2$ possibilities, we consider the combination that naturally arises from $R_{MNPQ}(\Omega_+)^2$.  In particular, using both the transpose relations (\ref{eq:transprel}) and swapping relations (\ref{eq:swaprel}), we can obtain
\begin{align}
    \Tr(\nabla^\mu\nabla^\nu\mathcal  S\nabla_\mu\nabla_\nu\mathcal S)\kern-4em&\nonumber\\
    &=-2\Tr(\nabla^\mu N^\nu_-\nabla_\mu N_{\nu+})-\Tr((\nabla^\mu N^\nu_-)(N_{\mu-}N_{\nu+}+N_{\mu_+}N_{\nu-}-2N_{\mu+}N_{\nu+}))\nn\\
    &\quad+\ft12\Tr(N^\mu_-N^\nu_+N_{\mu-}N_{\nu+}+N^\mu_-N^\nu_-N_{\mu+}N_{\nu+}+2N^\mu_-N^\nu_+N_{\mu+}N_{\nu+})\nn\\
    &\quad+\ft12\Tr(N^\mu_+N_{\mu-}N^\nu_+N_{\nu-}+N^\mu_-N_{\mu+}N^\nu_-N_{\nu+}-2N^\mu_-N_{\mu+}N^\nu_+N_{\nu-}).
\label{eq:4di2}
\end{align}
This expression allows us to rewrite the $\Tr(\nabla N\nabla N)$ that shows up in $R_{MNPQ}(\Omega_+)^2$ in terms of the T-duality invariant $\Tr(\nabla\nabla\mathcal S\nabla\nabla\mathcal S)$ along with higher-point terms of the form $(\nabla N)N^2$ and $N^4$.

Following the rewriting of $(\nabla N)^2$, we would then turn to terms of the form $(\nabla N)N^2$. Since there are an odd number of $N_\pm$ matrices, it cannot have alternating signs.  More problematically, however, it cannot appear from any combination of $S$ and its derivatives. Superficially, it seems it can come from an invariant of the form $\Tr((\nabla\nabla\mathcal S)(\partial\mathcal S)^2)$. However, we can check from (\ref{eq:Scovars}) that
\begin{equation}
    \Tr((\nabla\nabla\mathcal S)(\partial\mathcal S)^2)\sim\Tr(N_+N_-N_+N_-).
\end{equation}
While the right-hand side has the proper alternating combination of $N_+$ and $N_-$ for $O(d,d)$ invariance, it does not have the form of $\Tr((\nabla N)N^2)$.  This can also be seen more abstractly from (\ref{eq:Scovars}) where the $\nabla N$ terms enter off-diagonally in $\nabla\nabla\mathcal S$ while the $N^2$ terms enter diagonally in $(\partial\mathcal S)^2$.  The general pattern, which persists beyond the four-derivative level, is that the combination $\nabla N+Y$ can only enter in even powers regardless of whether there are an even or odd number of $\nabla\nabla\mathcal S$ factors in the trace.  Thus, at the lowest order in the $n$-point function expansion, we have the even/odd split
\begin{align}
    (\nabla\nabla\mathcal S)^{2k}(\partial\mathcal S)^{2l}&\sim N^{2l}(\nabla N)^{2k}+\cdots,\nn\\
    (\nabla\nabla\mathcal S)^{2k+1}(\partial\mathcal S)^{2l}&\sim N^{2l+2}(\nabla N)^{2k}+\cdots.
\end{align}
In both cases, the right-hand side starts as an even-point function.

The consequence of this is that, once the $\Tr(\nabla N\nabla N)$ terms are rewritten in terms of the $\Tr(\nabla\nabla\mathcal S\nabla\nabla\mathcal S)$ invariant, the remaining $\Tr((\nabla N)N^2)$ terms cannot form an invariant, and must be dealt with using integration by parts and the equations of motion.  This is an example where field redefinitions are required in order to rewrite the higher-derivative couplings in a more canonical form.

Note that, for the cosmological reduction, we can obtain two independent invariants
\begin{align}
    \Tr\relax(\ddot{\mathcal S}^2)&=-2\Tr\relax(\dot N_+\dot N_-)-\Tr\relax(\dot N_-N_-N_+)-\Tr\relax(\dot N_+N_+N_-)+\Tr\relax(\dot N_-N_+^2)+\Tr\relax(\dot N_+N_-^2)\nn\\
    &\qquad+2\Tr(N_+N_-N_+N_-)-\Tr(N_+^2N_-^2)+\ft12\Tr(N_+N_-^3)+\ft12\Tr(N_-N_+^3),\nn\\
    \Tr\relax(\dot{\mathcal S}^4)&=2\Tr(N_+N_-N_+N_-).
\label{eq:4di}
\end{align}
These correspond to (\ref{eq:4di1}) and (\ref{eq:4di2}).

\subsubsection{Field redefinitions}

As indicated above, we have to handle terms such as $\Tr((\nabla N)N^2)$ that do not directly correspond to any T-duality invariants using integration by parts and the lower-order equations of motion.  This amounts to performing a field redefinition. To motivate how this arises, we first write down the lower-dimensional equations of motion for the scalar sector coupled to gravity.  In particular, ignoring KK and winding gauge fields and the antisymmetric tensor, we have
\begin{align}
    0&=\mathcal E_{\mu\nu} \equiv R_{\mu\nu}+\nabla_\mu\nabla_\nu\Phi-\ft14\Tr(N_{(\mu+}N_{\nu)-}),\nn\\
    0&=\kern.9em\mathcal E\equiv R+2\Box\Phi-(\partial\Phi)^2-\ft14\Tr(N^\mu_+N_{\mu-}),\nn\\
    0&=\kern.4em\mathcal E_\pm\equiv\nabla^\mu N_{\mu\pm}-\partial^\mu\Phi N_{\mu\pm}\mp\ft12(N^\mu_+-N^\mu_-)N_{\mu\pm}.
\end{align}
We now consider the total derivative
\begin{align}
    \nabla^\mu[e^{-\Phi}\Tr(N_{\mu \tilde a}N^\nu_{\tilde b}N_{\nu \tilde c})]&\nn\\
    &\kern-4em=\Tr(\nabla^\mu(e^{-\Phi}N_{\mu \tilde a})N^\nu_{\tilde b} N_{\nu \tilde c})+e^{-\Phi}\Tr[(\nabla^\mu N^\nu_{\tilde b})N_{\nu \tilde c}N_{\mu \tilde a}+(\nabla^\mu N_{\nu \tilde c})N_{\mu \tilde a}N^\nu_{\tilde b}],
\end{align}
where $\tilde a,$ $\tilde b$ and $\tilde c$ denote $\pm$ (and should not be confused with the $T^{d}$ indices).  We can use the scalar equation of motion $\mathcal E_\pm$ to rewrite this as
\begin{align}
    \Tr[(\nabla^\mu N^\nu_{ \tilde b})N_{\nu \tilde c}N_{\mu \tilde a}+(\nabla^\mu N_{\nu \tilde c})N_{\mu \tilde a}N^\nu_{\tilde b}]&=e^{\Phi} \nabla^\mu[e^{-\Phi}\Tr(N_{\mu \tilde a}N^\nu_{\tilde b} N_{\nu \tilde c})]-\Tr(\mathcal E_{\tilde a} N^\nu_{\tilde b}N_{\nu \tilde c})\nn\\
    &\qquad-a\Tr(M^\mu N_{\mu \tilde a}N^\nu_{\tilde b}N_{\nu \tilde c}).
\end{align}
The left-hand side is a sum of two terms of the form $(\nabla N)N^2$.  However, they can be disentangled by making use of the transpose and swapping relations and taking appropriate linear combinations of the resulting expressions.  The first term on the right-hand side is a total derivative when used in the tree-level four-derivative action, while the second term can be removed by a field redefinition.  This leaves the final term, so we may write
\begin{equation}
     \Tr[(\nabla^\mu N^\nu_{\tilde b})N_{\nu \tilde c}N_{\mu \tilde a}+(\nabla^\mu N_{\nu \tilde c})N_{\mu \tilde a}N^\nu_{\tilde b}]~\to~-\tilde a\Tr(M^\mu N_{\mu \tilde a}N^\nu_{\tilde b} N_{\nu \tilde c}),
\label{eq:ibpi}
\end{equation}
where $\to$ indicates equivalence up to a total derivative and field redefinition. In general, we can write down $2\cdot2^3=16$ possible terms of the form $\Tr((\nabla N)NN)$, where the first 2 comes from the choice of index structure being either $\mu\nu\mu\nu$ or $\mu\nu\nu\mu$ and the $2^3$ comes from the $\pm$ choices on the three $N$'s. Note that we ignore the $\mu\mu\nu\nu$ index structure as that can be reduced by the $\mathcal E_\pm$ equation of motion. However, we can reduce these 16 possible combinations to a set of only four linearly independent expressions by use of transposing via \eqref{eq:transprel} and index swapping via \eqref{eq:swaprel}.  In particular, by transposing, we can choose to write $(\nabla^\mu N^\nu_-)NN$ and keep only the $\nabla N$ term with a minus sign. In addition, by index swapping, we can reduce to the $\mu\nu\mu\nu$ index structure.  We are thus left with four independent terms of the form $\Tr((\nabla^\mu N^\nu_{-})N_{\mu a}N_{\nu b})$.

We choose $\tilde b$ and $\tilde c$ to be various combinations of $\pm$ in (\ref{eq:ibpi}) and canonicalize the resulting expressions. This lets us obtain
\begin{align}
    2\Tr((\nabla^\mu N^\nu_-)N_{\mu -}N_{\nu -})&~\to~\ft12\Tr\relax[(N^\mu_+N_{\mu-}+N^\mu_-N_{\mu+})N^\nu_-N_{\nu-}\nn\\
    &\kern4em-(N^\mu_+N^\nu_-+N^\mu_-N^\nu_-)N_{\mu-}N_{\nu-}],\nn\\
    \Tr((\nabla^\mu N^\nu_-)(N_{\mu+}N_{\nu -}+N_{\mu -}N_{\nu +}))&~\to~\ft12\Tr\relax[(N^\mu_+N_{\mu-}+N^\mu_-N_{\mu+})N^\nu_-N_{\nu-}\nn\\
    &\kern4em-(N^\mu_+N^\nu_-+N^\mu_-N^\nu_-)N_{\mu-}N_{\nu+}],\nn\\
    \Tr((\nabla^\mu N^\nu_-)(N_{\mu+}N_{\nu+}+N_{\mu -}N_{\nu +}))&~\to~\ft12\Tr\relax[(N^\mu_+N_{\mu-}+N^\mu_-N_{\mu+})N^\nu_+N_{\nu-}\nn\\
    &\kern4em-(N^\mu_+N^\nu_-+N^\mu_-N^\nu_-)N_{\mu+}N_{\nu+}],\nn\\
    \Tr((\nabla^\mu N^\nu_-)(N_{\mu+}N_{\nu+}+N_{\mu+}N_{\nu -}))&~\to~\ft12\Tr\relax[(N^\mu_+N_{\mu-}+N^\mu_-N_{\mu+})N^\nu_-N_{\nu+}\nn\\
    &\kern4em-(N^\mu_+N^\nu_-+N^\mu_-N^\nu_-)N_{\mu+}N_{\nu-}].
\end{align}
Finally, we can take linear combinations of the last three expressions to obtain
\begin{align} \label{eq: field redefinitions}
    2\Tr((\nabla^\mu N^\nu_-)N_{\mu+}N_{\nu+})&~\to~\ft12\Tr\relax[(N^\mu_+N_{\mu-}+N^\mu_-N_{\mu+})(N^\nu_+N_{\nu-}+N^\nu_-N_{\nu+}-N^\nu_-N_{\nu-})\nn\\
    &\kern4em-(N^\mu_+N^\nu_-+N^\mu_-N^\nu_-)(N_{\mu+}N_{\nu+}+N_{\mu+}N_{\nu-}-N_{\mu-}N_{\nu+})],\nn\\
    2\Tr((\nabla^\mu N^\nu_-)N_{\mu+}N_{\nu-})&~\to~\ft12\Tr\relax[(N^\mu_+N_{\mu-}+N^\mu_-N_{\mu+})(-N^\nu_+N_{\nu-}+N^\nu_-N_{\nu+}+N^\nu_-N_{\nu-})\nn\\
    &\kern4em-(N^\mu_+N^\nu_-+N^\mu_-N^\nu_-)(-N_{\mu+}N_{\nu+}+N_{\mu+}N_{\nu-}+N_{\mu-}N_{\nu+})],\nn\\
    2\Tr((\nabla^\mu N^\nu_-)N_{\mu-}N_{\nu+})&~\to~\ft12\Tr\relax[(N^\mu_+N_{\mu-}+N^\mu_-N_{\mu+})(N^\nu_+N_{\nu-}-N^\nu_-N_{\nu+}+N^\nu_-N_{\nu-})\nn\\
    &\kern4em-(N^\mu_+N^\nu_-+N^\mu_-N^\nu_-)(N_{\mu+}N_{\nu+}-N_{\mu+}N_{\nu-}+N_{\mu-}N_{\nu+})].
\end{align}
We are now ready to revisit the four-derivative couplings in the heterotic and bosonic strings, making use of these field redefinitions.

\subsection{The heterotic string at $\mathcal{O}(\alpha')$}
The gravitational sector of the heterotic string action at the four-derivative level \cite{Metsaev:1987zx,Bergshoeff:1988nn,Bergshoeff:1989de,Chemissany:2007he} is given by
\begin{align}
S_{H}=\int d^{10} x \sqrt{-g} e^{-2 \phi}\left(R+4(\partial \phi)^{2}-\frac{1}{12} \widehat{H}_{ABC}\widehat{H}^{ABC}+\fft18\alpha'R_{ABCD}(\Omega_+)R^{ABCD}(\Omega_+)\right),
\label{eq:lagHet}
\end{align}
where we have ignored the heterotic gauge fields.  Here, the three-form field strength $\widehat H$ has a non-trivial Bianchi identity
\begin{equation}
    d\widehat H=-\fft14\alpha'\Tr R(\Omega_+)\wedge R(\Omega_+),
\end{equation}
corresponding to the addition of the Lorentz Chern-Simons term
\begin{equation}
    \widehat{H}=dB-\frac{\alpha^{\prime}}{4}\omega_{3L}(\Omega_{+}),
\label{eq:hetlcs}
\end{equation}
where
\begin{equation}
    \omega_{3L}(\Omega_+)=\Tr(\Omega_+\wedge d\Omega_++\fft23\Omega_+\wedge\Omega_+\wedge\Omega_+).
\label{Chern-Simons term}
\end{equation}
By expanding $\widehat{H}^2$ perturbatively up to $\mathcal O(\alpha')$, we see that the effective four-derivative Lagrangian at this level is given by
\begin{equation}
    e^{-1}\mathcal L_H^{\partial^4}=\fft18\alpha'e^{-2\phi}\left(R_{ABCD}(\Omega_+)R^{ABCD}(\Omega_+)+\fft13H_{ABC}\omega_{3L}^{ABC}(\Omega_+)\right).
\label{eq:het4d}
\end{equation}
There are two terms to reduce on the torus.  For the Riemann-squared term, the direct reduction (\ref{eq:Riemann}) gives
\begin{align}
    R_{ABCD}(\Omega_+)R^{ABCD}(\Omega_+)\kern-8em&\nn\\
    &=R_{\mu\nu\rho\sigma}(\omega_+)R^{\mu\nu\rho\sigma}(\omega_+)+\Tr\relax((\nabla^{\mu\,(+)} N^\nu_-)g^{-1}(\nabla_\mu^{(+)} N_{\nu-})^Tg\relax)\nn\\&\quad+\Tr\relax((\nabla^{\mu\,(+)} N^\nu_-)N_{\nu+}N_{\mu-}\relax)
    +\fft18\Tr(N^{\mu}_{-}N^{\nu}_{+})\Tr(N_{\mu -}N_{\nu +})-\fft38\Tr(N^\mu_+N^\nu_-N_{\mu+}N_{\nu-})\nn\\
    &\quad+\fft38\Tr(N^\mu_-N_{\mu+}N^\nu_-N_{\nu+})+\fft18\Tr(N^\mu_+N_{\mu-}N^\nu_+N_{\nu-}).
\end{align}
After rewriting $\nabla_\mu^{(+)}$ in terms of $\nabla_\mu$ and $h_{\mu\nu\rho}$ using (\ref{eq:nabla+}) and using the transpose relation, (\ref{eq:transprel}), we end up with the second derivative terms in the canonical form $\Tr(\nabla^\mu N^\nu_-\nabla_\mu N_{\nu+})$.  This in turn can be replaced by $\Tr(\nabla\nabla\mathcal S\nabla\nabla\mathcal S)$ using (\ref{eq:4di2}).  The result is
\begin{align}
        R_{ABCD}(\Omega_{+})R^{ABCD}(\Omega_{+})\kern-8em&\nn\\
        &=R_{\mu\nu\rho\sigma}(\omega_{+})R^{\mu\nu\rho\sigma}(\omega_{+})+\fft14h^{\mu\rho\sigma}h^\nu{}_{\rho\sigma}\Tr(N_{\mu+}N_{\nu-})-\fft12h^{\mu\nu\rho}\Tr(N_{\mu+}N_{\nu-}N_{\rho-})\nn\\ &\quad-\frac{1}{2}\Tr(\nabla^{\mu}\nabla^{\nu}\mathcal S\nabla_{\mu}\nabla_{\nu}\mathcal S)+\Tr((\nabla^\mu N^\nu_-)N_{\mu+}N_{\nu+})\nn\\
        &\quad+ \frac{1}{8}\Tr(N^{\mu}_{-}N^{\nu}_{+})\Tr(N_{\mu -}N_{\nu +})+\frac{1}{8}\Tr(N^{\mu}_{+}N^{\nu}_{-}N_{\mu +}N_{\nu -}) +\frac{1}{2}\Tr(N^{\mu}_{+}N^{\nu}_{-}N_{\mu-}N_{\nu-})\nn\\
        &\quad + \frac{3}{8}\Tr(N^{\mu}_{+}N_{\mu-}N^{\nu}_{+}N_{\nu-}) + \frac{3}{8}\Tr(N^{\mu}_{-}N_{\mu +}N^{\nu}_{-}N_{\nu +})
        -\frac{1}{4}\Tr(N^{\mu}_{-}N_{\mu +}N^{\nu}_{+}N_{\nu-}).
\end{align}
After writing the two-point contribution as $\Tr(\nabla\nabla\mathcal S\nabla\nabla\mathcal S)$, there is still a three-point contribution of the form $\Tr((\nabla N)N^2)$ left over.  As discussed above, this does not form an $O(d,d)$ invariant, but can be removed by a field redefinition as given by the first line of (\ref{eq: field redefinitions}).  After doing so, we arrive at the result
\begin{align}
    R_{ABCD}(\Omega_{+})R^{ABCD}(\Omega_{+})\kern-8em &\nn\\
    &\to R_{\mu\nu\rho\sigma}(\omega_{+})R^{\mu\nu\rho\sigma}(\omega_{+}) +\fft14h^{\mu\rho\sigma}h^\nu{}_{\rho\sigma}\Tr(N_{\mu+}N_{\nu-})\nn\\
    &\quad-\frac{1}{2}\Tr(\nabla^{\mu}\nabla^{\nu}\mathcal S\nabla_{\mu}\nabla_{\nu}\mathcal S) + \frac{1}{8}\Tr(N^{\mu}_{-}N^{\nu}_{+})\Tr(N_{\mu -}N_{\nu +}) \nn\\
    &\quad + \frac{5}{8}\Tr(N^{\mu}_{+}N_{\mu-}N^{\nu}_{+}N_{\nu-}) + \frac{5}{8}\Tr(N^{\mu}_{-}N_{\mu +}N^{\nu}_{-}N_{\nu +})-\frac{1}{8}\Tr(N^{\mu}_{+}N^{\nu}_{-}N_{\mu +}N_{\nu -})\nn\\
    &\quad +\frac{1}{4}\Tr(N^{\mu}_{-}N_{\mu +}N^{\nu}_{+}N_{\nu-}) - \frac{1}{4}\Tr(N^{\mu}_{+}N_{\mu -}N^{\nu}_{-}N_{\nu-})-\frac{1}{4}\Tr(N^{\mu}_{-}N_{\mu +}N^{\nu}_{-}N_{\nu-})\nn\\
    &\quad +\frac{1}{4}\Tr(N^{\mu}_{+}N^{\nu}_{-}N_{\mu-}N_{\nu-})-\fft12h^{\mu\nu\rho}\Tr(N_{\mu+}N_{\nu-}N_{\rho-}).
\label{eq: Riemann squared invariant}
\end{align}
We have written the terms such that the first three lines on the right hand side are explicitly T-dual invariant while the remaining two lines are not.  This demonstrates that the Riemann-squared term by itself is not a good invariant.

To complete the heterotic four-derivative invariant, we also need the second term in (\ref{eq:het4d}).  Using the torus reduction of the torsionful spin connection, (\ref{eq:Omegapm}), we can obtain the reduction of the Lorentz Chern-Simons term
\begin{align}
    \omega_{3L\,\mu\nu\rho}(\Omega_+)&=\omega_{3L\,\mu\nu\rho}(\omega_+)-3\Tr(e^{-1}\partial_{[\mu} ee^{-1}\partial_\nu ee^{-1}\partial_{\rho]} e)+3\partial_{[\mu}\Tr(e^{-1}\partial_\nu eM_{\rho]})\nn\\
    &\qquad+\fft14\Tr(3N_{[\mu+}N_{\nu-}N_{\rho]-}+N_{[\mu-}N_{\nu-}N_{\rho]-})\nn\\
    \omega_{3L\,\mu}{}^i{}_j(\Omega_+)&=\left[N^\nu_+\nabla_\mu^{(+)}N_{\nu-}+\fft12N^\nu_+N_{\mu+}N_{\nu+}\right]^i{}_j.
\end{align}
Contracting with $H_{MNP}$ then gives
\begin{align}
    \begin{split}
    H_{MNP}\omega_{3L}^{MNP}(\Omega_+) &= h^{\mu\nu\rho}\biggl(\omega_{3L\,\mu\nu\rho}(\omega_+)-3\Tr(e^{-1}\partial_{[\mu} ee^{-1}\partial_\nu ee^{-1}\partial_{\rho]} e)+3\partial_{[\mu}\Tr(e^{-1}\partial_\nu eM_{\rho]})   \nn\\ &\kern4em+\fft14\Tr(9N_{[\mu+}N_{\nu-}N_{\rho]-}+N_{[\mu-}N_{\nu-}N_{\rho]-})\biggr)\nn\\
    &\quad-\frac{3}{2}\Tr((\nabla^{\mu} N^{\nu}_-)N_{\mu+}N_{\nu +}) + \frac{3}{2}\Tr((\nabla^{\mu} N^{\nu}_-)N_{\mu-}N_{\nu+}) \\
    &\quad - \frac{3}{4} \Tr(N^{\mu}_+N^{\nu}_+N_{\mu+}N_{\nu-})+ \frac{3}{4}\Tr(N^{\mu}_-N^{\nu}_+N_{\mu+}N_{\nu-}).
    \end{split}
\end{align}
As mentioned above, field redefinitions are required to write the $\Tr((\nabla N)N^2)$ terms in a more canonical way.  Using \eqref{eq: field redefinitions}, we find
\begin{align}
    H_{MNP}\omega_{3L}^{MNP}(\Omega_+)\kern-6em& \nn\\
    &\to 
    h^{\mu\nu\rho}\biggl(\omega_{3L\,\mu\nu\rho}(\omega_+)-3\Tr(e^{-1}\partial_{[\mu} ee^{-1}\partial_\nu ee^{-1}\partial_{\rho]} e)+3\partial_{[\mu}\Tr(e^{-1}\partial_\nu eM_{\rho]})   \nn\\ &\kern4em+\fft14\Tr(9N_{[\mu+}N_{\nu-}N_{\rho]-}+N_{[\mu-}N_{\nu-}N_{\rho]-})\biggr)\nn\\
    &\quad+\frac{3}{4} \Tr(N^{\mu}_{+}N^{\nu}_{-}N_{\mu +}N_{\nu -})
    - \frac{3}{4}\Tr(N^{\mu}_{-}N_{\mu+}N^{\nu}_{-}N_{\nu+})
    +\frac{3}{4}\Tr(N^{\mu}_{-}N_{\mu +}N^{\nu}_{-}N_{\nu-})\nn\\
    & \quad - \frac{3}{4}\Tr(N^{\mu}_{-}N_{\mu+}N^{\nu}_{+}N_{\nu-})
    -\frac{3}{4}\Tr(N^{\mu}_{+}N^{\nu}_{+}N_{\mu +}N_{\nu -})  +\frac{3}{4}\Tr(N^{\mu}_{+}N_{\mu -}N^{\nu}_{-}N_{\nu-}).
\label{eq:Hw3L}
\end{align}
A quick inspection of this expression shows that the first two $\Tr(N^4)$ terms are in the T-dual invariant form of alternating signs of $N$ while the rest are not.

Combining (\ref{eq: Riemann squared invariant}) and (\ref{eq:Hw3L}), we can now see that the reduced four-derivative Lagrangian is given by
\begin{align}
        e^{-1}\mathcal{L}^{\partial^{4}}_{H}&=
        \frac{\alpha'}{8}e^{-\Phi}\biggl[R_{\mu\nu\rho\sigma}(\omega_+)^2+\fft13h^{\mu\nu\rho}W_{\mu\nu\rho}+\fft14h^{\mu\rho\sigma}h^\nu{}_{\rho\sigma}\Tr(N_{\mu+}N_{\nu-})\nn\\ &\quad-\frac{1}{2}\Tr(\nabla^{\mu}\nabla^{\nu}\mathcal S\nabla_{\mu}\nabla_{\nu}\mathcal S) + \frac{1}{8}\Tr(N^{\mu}_{-}N^{\nu}_{+})\Tr(N_{\mu-}N_{\nu+}) \nn\\
        &\quad + \frac{5}{8}\Tr(N^{\mu}_{+}N_{\mu-}N^{\nu_+}N_{\nu-})+ \frac{3}{8}\Tr(N^{\mu}_{-}N_{\mu+}N^{\nu}_{-}N_{\nu+})+\frac{1}{8}\Tr(N^{\mu}_{+}N^{\nu}_{-}N_{\mu +}N_{\nu -})\biggr],
\end{align}
where we have defined the shifted Chern-Simons term
\begin{align}
    W_{\mu\nu\rho}&=\omega_{3L\,\mu\nu\rho}(\omega_+)-3\Tr(e^{-1}\partial_{[\mu} ee^{-1}\partial_\nu ee^{-1}\partial_{\rho]} e)+3\partial_{[\mu}\Tr(e^{-1}\partial_\nu eM_{\rho]})\nn\\ &\quad+\fft14\Tr(3N_{[\mu+}N_{\nu-}N_{\rho]-}+N_{[\mu-}N_{\nu-}N_{\rho]-}).
\label{eq:Wterm}
\end{align}
Note that this is manifestly $O(d,d)$ invariant, except for the $h^{\mu\nu\rho}W_{\mu\nu\rho}$ term, as can be seen from the structure of the alternating signs. Written in terms of the scalar matrix $\mathcal S$, we find
\begin{align}
        e^{-1}\mathcal{L}^{\partial^{4}}_{H}&=
        \frac{\alpha'}{8}e^{-\Phi}\biggl[R_{\mu\nu\rho\sigma}(\omega_+)^2+\fft13h^{\mu\nu\rho}W_{\mu\nu\rho}-\fft18h^{\mu\rho\sigma}h^\nu{}_{\rho\sigma}\Tr(\partial_\mu\mathcal S\partial_\nu\mathcal S)\nn\\ &\kern5em-\frac{1}{2}\Tr(\nabla^{\mu}\nabla^{\nu}\mathcal S\nabla_{\mu}\nabla_{\nu}\mathcal S) + \frac{1}{32}\Tr(\nabla_{\mu}\mathcal S\nabla_{\nu}\mathcal S)\Tr(\nabla^{\mu}\mathcal S\nabla^{\nu}\mathcal S) \nn\\
        &\kern5em + \frac{1}{2}\Tr(\nabla^\mu\mathcal S\nabla_\mu\mathcal S\nabla^\nu\mathcal S\nabla_\nu\mathcal S) - \frac{1}{8}\Tr(\mathcal S\nabla^\mu\mathcal S\nabla_\mu\mathcal S\nabla^\nu\mathcal S\nabla_\nu\mathcal S)\nn\\
        &\kern5em+\frac{1}{16}\Tr(\nabla_\mu\mathcal S\nabla_\nu\mathcal S\nabla^\mu\mathcal S\nabla^\nu\mathcal S)\biggr],
\label{eq: heterotic lagrangian alpha prime}
\end{align}
where we have used \eqref{eq:4di1}.

Turning now to the $h^{\mu\nu\rho}W_{\mu\nu\rho}$ term, we see that the final term in (\ref{eq:Wterm}) breaks $O(d,d)$ invariance.  However, as noted in \cite{Eloy:2020dko}, $O(d,d)$ can be restored by a Green-Schwarz like mechanism where $W_{\mu\nu\rho}$ is absorbed into a shifted $h$-field
\begin{equation}
    \hat h_{\mu\nu\rho}=3\partial_{[\mu}b_{\nu\rho]}-\fft{\alpha'}4W_{\mu\nu\rho},
\end{equation}
in analogy with the original ten-dimensional expression (\ref{eq:hetlcs}).  Moreover, the WZW term in (\ref{eq:Wterm}) as well as the following term are locally total derivatives and can be removed by shifting $b_{\mu\nu}$.  After doing so, we have explicitly
\begin{equation}
    \hat h_{\mu\nu\rho}=3\partial_{[\mu}\hat b_{\nu\rho]}-\fft{\alpha'}4\left(\omega_{3L\,\mu\nu\rho}(\omega_+)+\omega_{3N\,\mu\nu\rho}\right),
\label{eq:hath}
\end{equation}
where
\begin{equation}
    \omega_{3N\,\mu\nu\rho}=\fft14\Tr(3N^{\mu}_{+}N^{\nu}_{-}N^{\rho}_{-} +N^{\mu}_{-}N^{\nu}_{-}N^{\rho}_{-}).
\label{eq:omega3N}
\end{equation}
The shifted $\hat b_{\mu\nu}$ then transforms non-trivially under $O(d,d)$ to compensate for the transformation of $\omega_{3N\,\mu\nu\rho}$ \cite{Eloy:2020dko}.

\subsection{The bosonic string at $\mathcal{O}(\alpha')$}
We now examine the bosonic string at $\mathcal{O}(\alpha')$ whose action at this order is \cite{Metsaev:1987zx}
\begin{align}
\int d^{D} x \sqrt{-g} e^{-2 \phi}\fft14\alpha'\left(R_{ABCD}R^{ABCD}-\frac{1}{2} H^{ABE} H^{CD}{ }_{E} R_{ABCD}+\frac{1}{24} H^{4}-\frac{1}{8}\left(H_{AB}^{2}\right)^{2}\right),
\end{align}
where we have defined $H^4$ and $H^{2}_{AB}$ as
\begin{align}
\begin{aligned}
H^{4} & 
\equiv H_{ABC} \tensor{H}{^{A}_{D}^{E}} \tensor{H}{^{B}_{E}^{F}} \tensor{H}{^{C}_{F}^{D}}, \qquad
H_{AB}^{2} \equiv \tensor{H}{_{A}^{CD}} H_{B CD}, \qquad
\left(H_{AB}^{2}\right)^{2} \equiv H_{AB}^{2} H^{2 AB}, 
\end{aligned}
\end{align}
We convert the torsion free Riemann curvature to Riemann curvature with torsion which allows us to make contact with the $N$ matrices that appear in the T-dual invariants. The Langrangian can then be rewritten as \cite{David:2021jqn}
%
\begin{equation}
    e^{-1}\mathcal{L}_{B}^{\partial^4} = \fft14\alpha'e^{-2\phi}\left(R_{MNPQ}(\Omega_+)^{2} -  R_{MNPQ}(\Omega_+)H^{MNR}\tensor{H}{^{PQ}_{R}} -\frac{1}{3} H^{4}\right).
\label{eq:bapm}
\end{equation}
We now follow a similar procedure as for the case of the heterotic string using the methodology outlined in subsection~\ref{subsection: General approach to verifying T-duality}. We find
\begin{align}
    \begin{split} \label{eq: H4}
        H^4 &= h^4 + h^{\mu\nu\rho}\Tr(3N_{\mu+}N_{\nu-}N_{\rho-}-N_{\mu-}N_{\nu-}N_{\rho-}) + \frac{3}{8}\Tr(N^{\mu}_{-}N^{\nu}_{-}N_{\mu-}N_{\nu-})\\&\quad-\frac{3}{2}\Tr (N^{\mu}_{+}N^{\nu}_{-}N_{\mu-}N_{\nu-})+\frac{3}{8}\Tr(N^{\mu}_{+}N^{\nu}_{-}N_{\mu+}N_{\nu-})+\frac{3}{4}\Tr(N^{\mu}_{+}N^{\nu}_{+}N_{\mu-}N_{\nu-}).
    \end{split}
\end{align}
Note that we did not need field redefinitions as there were no terms involving $\nabla N$. On the other hand, $R(\Omega_+)H^2$ requires us to utilize the field redefinitions \eqref{eq: field redefinitions} such that
\begin{align}
    \begin{split} \label{eq: RHsquared}
        R_{MNPQ}(\Omega_+)H^{MNR}H^{PQ}{}_R \kern-8em\\
        &\to R_{\mu\nu\rho\sigma}h^{\mu\nu\lambda}h^{\rho\sigma}{}_\lambda - \fft14h_{\mu\nu\rho}\Tr(7N^{\mu}_{+}N^{\nu}_{-}N^{\rho}_{-}-N^{\mu}_{-}N^{\nu}_{-}N^{\rho}_{-})\\
        &\quad+\frac{1}{8}\Tr(N^{\mu}_{+}N_{\mu-}N^\nu_{+}N_{\nu-})+\frac{1}{8}\Tr(N^{\mu}_{-}N_{\mu+}N^\nu_-N_{\nu+})\\
        &\quad+\frac{1}{4}\Tr(N^{\mu}_{+}N_{\mu-}N^{\nu}_-N_{\nu+})-\frac{1}{4}\Tr(N^{\mu}_{+}N_{\mu-}N^\nu_-N_{\nu-})-\frac{1}{4}\Tr(N^{\mu}_{-}N_{\mu+}N^{\nu}_-N_{\nu-}) \\
        &\quad-\frac{3}{8}\Tr(N^{\mu}_{+}N^{\nu}_{-}N_{\mu+}N_{\nu-})-\frac{1}{4}\Tr(N^{\mu}_{+}N^{\nu}_{+}N_{\mu-}N_{\nu-})+\frac{3}{4}\Tr(N^{\mu}_{+}N^{\nu}_{-}N_{\mu-}N_{\nu-})\\&\quad-\frac{1}{8}\Tr(N^{\mu}_{-}N^{\nu}_{-}N_{\mu-}N_{\nu-}).
    \end{split}
\end{align}
Combining \eqref{eq: Riemann squared invariant}, \eqref{eq: H4} and \eqref{eq: RHsquared}, we find
\begin{align}
        e^{-1}\mathcal{L}_{B}^{\partial^4} &=\fft14\alpha'e^{-\Phi}\biggl[ R_{\mu\nu\rho\sigma}(\omega_+)^2-R_{\mu\nu\rho\sigma}(\omega_+)h^{\mu\nu\lambda}h^{\rho\sigma}{}_\lambda-\fft13h^4-\fft18h^{\mu\rho\sigma}h^\nu{}_{\rho\sigma}\Tr(\partial_\mu\mathcal S\partial_\nu\mathcal S)\nn \\
        &\kern5em+\frac{1}{3} h_{\mu\nu\rho}\omega_{3N}^{\mu\nu\rho}- \frac{1}{2}\Tr(\nabla^{\mu}\nabla^{\nu}\mathcal S\nabla_{\mu}\nabla_{\nu}\mathcal S) + \frac{1}{32}\Tr(\nabla^{\mu}\mathcal S\nabla^{\nu}\mathcal S)\Tr(\nabla_{\mu}\mathcal S\nabla_{\nu}\mathcal S)\nn \\
        &\kern5em + \frac{1}{2}\Tr(\nabla^{\mu}\mathcal S\nabla_{\mu}\mathcal S\nabla^{\nu}\mathcal S\nabla_{\nu}\mathcal S) + \frac{1}{16}\Tr(\nabla^{\mu}\mathcal S\nabla^{\nu}\mathcal S\nabla_{\mu}\mathcal S\nabla_{\nu}\mathcal S)\biggr],
\label{eq:bosred}
\end{align}
where $\omega_{3N}^{\mu\nu\rho}$ is defined in (\ref{eq:omega3N}).  This expression is explicitly $O(d,d)$ invariant except for the $h_{\mu\nu\rho}\omega_{3N}^{\mu\nu\rho}$ term.  However it can be treated just as in the heterotic case by defining a shifted $h$-field \cite{Eloy:2020dko}
\begin{equation}
    \tilde h_{\mu\nu\rho}=3\partial_{[\mu}b_{\nu\rho]}-\fft{\alpha'}2\omega^{3N}_{\mu\nu\rho}.
\end{equation}
This torus reduced Lagrangian, (\ref{eq:bosred}), agrees with \cite{Godazgar:2013bja} as well as \cite{Eloy:2020dko} after appropriate field redefinitions.

Note that the bosonic string correction, (\ref{eq:bosred}), does not precisely match the heterotic string correction, \eqref{eq: heterotic lagrangian alpha prime}.  The differences are that the heterotic case has  an additional $\Tr(\mathcal S\nabla^\mu\mathcal S\nabla_\mu\mathcal S\nabla^\nu\mathcal S\nabla_\nu\mathcal S)$ coupling in \eqref{eq: heterotic lagrangian alpha prime} as well as the lower-dimensional Lorentz Chern-Simons term $\omega_{3L}$ in (\ref{eq:hath}).  Since this additional term vanishes in the cosmological reduction, in that case there is no distinction between the reduced heterotic and bosonic string cases.

\section{The type II string and contact terms at $\mathcal{O}(\alpha'^3)$} \label{section: alpha prime cubed corrections}

As we have shown so far, the scalar $O(d,d)$ invariants are built out of the traces of $N_{\pm}$, which are ultimately linear combinations of the internal metric and antisymmetric tensor field and their derivatives. At order $\alpha'$, we find that the building blocks composed of $N_{\pm}$ are a convenient way to study the structure of the higher derivative terms that ensure T-duality. In this section, we promote the use of the $N_{\pm}$ matrices to investigate the structure of higher derivative corrections to the tree-level $\mathcal O(\alpha'^3)$ couplings of the type II string.

The general strategy is as before, where we first write the torus compactified action using the $N_\pm$ building blocks.  We then look to replace $\nabla N$ and $N$ by $\nabla\nabla S$ and $\nabla S$ starting from the four-point contact terms and working up to the eight-point terms.  In practice, since only partial results are known at five and higher points, we focus mostly on the four-point terms.

For the type II string, the first correction arises at the eight-derivative level.  In the NSNS sector, the tree-level four-point couplings takes the form \cite{Gross:1986iv,Gross:1986mw}
\begin{align}
    \begin{gathered}
        e^{-1}\mathcal{L}_{R(\Omega_{+})^4} \sim
        \alpha'e^{-2\phi}\left[t_{8} t_{8} R\left(\Omega_{+}\right)^{4}-\frac{1}{4} \epsilon_{8} \epsilon_{8} R\left(\Omega_{+}\right)^{4}\right].
    \end{gathered}
\end{align}
While the $\epsilon_{8} \epsilon_{8} R\left(\Omega_{+}\right)^{4}$ term does not actually contribute to the four-point function, its presence can be deduced from the structure of the string scattering amplitude.  The explicit expression for $t_{8} t_{8} R(\Omega_+)^{4}$ is given by
\begin{align} \label{t8 t8 expanded}
    \begin{split}
    t_{8} t_{8} R(\Omega_+)^{4}&=
    24\tensor{R}{^{\mu_{6} \mu_{5}}_{\nu_{8} \nu_{7}}}(\Omega_{+}) \tensor{R}{^{\mu_{8} \mu_{7}}_{\nu_{6} \nu_{5}}}(\Omega_{+}) \tensor{R}{_{\mu_{5} \mu_{6}}^{\nu_{5} \nu_{6}}}(\Omega_{+}) \tensor{R}{_{\mu_{7} \mu_{8}}^{\nu_{7} \nu_{8}}}(\Omega_{+})
    \\& \quad +
    12\tensor{R}{^{\mu_{6} \mu_{5}}_{\nu_{6} \nu_{5}}} (\Omega_{+})\tensor{R}{^{\mu_{8} \mu_{7}}_{\nu_{8} \nu_{7}}}(\Omega_{+}) \tensor{R}{_{\mu_{5} \mu_{6}}^{\nu_{5} \nu_{6}}}(\Omega_{+}) \tensor{R}{_{\mu_{7} \mu_{8}}^{\nu_{7} \nu_{8}}}(\Omega_{+})
    \\ & \quad + 
     192\tensor{R}{^{\mu_{4} \mu_{5}}_{\nu_{4} \nu_{5}}}(\Omega_{+}) \tensor{R}{_{\mu_{3} \mu_{4}}^{\nu_{3} \nu_{4}}}(\Omega_{+}) \tensor{R}{_{\mu_{5} \mu_{6}}^{\nu_{5} \nu_{6}}}(\Omega_{+}) \tensor{R}{^{\mu_{6} \mu_{3}}_{\nu_{6} \nu_{3}}}(\Omega_{+})
    \\& \quad +
    384 \tensor{R}{^{\mu_{8} \mu_{3}}_{\nu_{8} \nu_{5}}}(\Omega_{+}) \tensor{R}{_{\mu_{3} \mu_{5}}_{ \nu_{6} \nu_{7}}}(\Omega_{+}) \tensor{R}{^{\mu_{5} \mu_{7}}^{\nu_{5} \nu_{6}}}(\Omega_{+}) \tensor{R}{_{\mu_{7} \mu_{8}}^{\nu_{7} \nu_{8}}}(\Omega_{+})
    \\ & \quad
    -48\tensor{R}{^{\mu_{6} \mu_{5}}_{\nu_{8} \nu_{3}}}(\Omega_{+}) \tensor{R}{^{\mu_{8} \mu_{7}}^{ \nu_{3} \nu_{4}}}(\Omega_{+}) \tensor{R}{_{\mu_{5} \mu_{6}}_{\nu_{4} \nu_{7}}}(\Omega_{+}) \tensor{R}{_{\mu_{7} \mu_{8}}^{\nu_{7} \nu_{8}}}(\Omega_{+})
    \\ & \quad
    -32\tensor{R}{^{\mu_{6} \mu_{5}}_{\nu_{8} \nu_{5}}}(\Omega_{+}) \tensor{R}{^{\mu_{8} \mu_{7}}_{ \nu_{6} \nu_{7}}}(\Omega_{+}) \tensor{R}{_{\mu_{5} \mu_{6}^{\nu_{5} \nu_{6}}}}(\Omega_{+}) \tensor{R}{_{\mu_{7} \mu_{8}}^{\nu_{7} \nu_{8}}}(\Omega_{+})
    \\ & \quad 
   -48 \tensor{R}{_{\mu_{8} \mu_{3}}^{\nu_{6} \nu_{5}}}(\Omega_{+}) \tensor{R}{^{ \mu_{3} \mu_{4}}^{\nu_{8} \nu_{7}}}(\Omega_{+}) \tensor{R}{_{\mu_{4} \mu_{7}}_{\nu_{5} \nu_{6}}}(\Omega_{+}) \tensor{R}{^{\mu_{7} \mu_{8}}_{\nu_{7} \nu_{8}}}(\Omega_{+})
    \\ & \quad
    -96 \tensor{R}{_{\mu_{8} \mu_{5}}^{\nu_{6} \nu_{5}}}(\Omega_{+}) \tensor{R}{_{ \mu_{6} \mu_{7}}^{\nu_{8} \nu_{7}}}(\Omega_{+}) \tensor{R}{^{\mu_{5} \mu_{6}}_{\nu_{5} \nu_{6}}}(\Omega_{+}) \tensor{R}{^{\mu_{7} \mu_{8}}_{\nu_{7} \nu_{8}}}(\Omega_{+}).
    \end{split}
\end{align}
Note that the structure of the $t_8t_8$ invariant ensures that this expression only contains Riemann tensors in various combinations, and no Ricci contractions.  On the other hand, $\epsilon_8\epsilon_8 R(\Omega_+)^4$ contains both Riemann and Ricci curvatures
\begin{equation}
\begin{split}
    \epsilon_8\epsilon_8 R(\Omega_+)^4&=-1536 R^{\mu_1 \mu_2 \mu_3 \mu_4}(\Omega_+) R_{\mu_3}{}^{\mu_5}{}_{\mu_1}{}^{\mu_6}(\Omega_+) R_{\mu_4}{}^{\mu_7}{}_{\mu_5}{}^{\mu_8}(\Omega_+) R_{\mu_6 \mu_8 \mu_2 \mu_7}(\Omega_+)\\
    &\quad- 1536 R^{\mu_1 \mu_2 \mu_3 \mu_4}(\Omega_+) R_{\mu_3 \mu_4}{}^{\mu_5 \mu_6}(\Omega_+) R_{\mu_5}{}^{\mu_7}{}_{\mu_1}{}^{\mu_8}(\Omega_+) R_{\mu_6 \mu_8 \mu_2 \mu_7}(\Omega_+)\\
    &\quad+ 768 R^{\mu_1 \mu_2 \mu_3 \mu_4}(\Omega_+) R_{\mu_3}{}^{\mu_5}{}_{\mu_1}{}^{\mu_6}(\Omega_+) R_{\mu_4}{}^{\mu_7}{}_{\mu_2}{}^{\mu_8}(\Omega_+) R_{\mu_6 \mu_8 \mu_5 \mu_7}(\Omega_+)\\
    &\quad+ 96 R^{\mu_1 \mu_2 \mu_3 \mu_4}(\Omega_+) R_{\mu_3 \mu_4}{}^{\mu_5 \mu_6}(\Omega_+) R_{\mu_5 \mu_6}{}^{\mu_7 \mu_8}(\Omega_+) R_{\mu_7 \mu_8 \mu_1 \mu_2}(\Omega_+)\\
    &\quad- 768 R^{\mu_1 \mu_2 \mu_3 \mu_4}(\Omega_+) R_{\mu_3 \mu_4 \mu_1}{}^{\mu_5}(\Omega_+) R_{\mu_5}{}^{\mu_6 \mu_7 \mu_8}(\Omega_+) R_{\mu_7 \mu_8 \mu_2 \mu_6}(\Omega_+)\\
    &\quad+ 48 R^{\mu_1 \mu_2 \mu_3 \mu_4}(\Omega_+) R_{\mu_3 \mu_4 \mu_1 \mu_2}(\Omega_+) R^{\mu_5 \mu_6 \mu_7 \mu_8}(\Omega_+) R_{\mu_7 \mu_8 \mu_5 \mu_6}(\Omega_+)+\cdots,
\end{split}
\label{eq:e8e8R4x}
\end{equation}
where the ellipses denote invariant combinations involving the Ricci terms.

At the level of the four-point function, we only need the linearized reduction of Riemann.  Examination of (\ref{eq:Riemann}) shows that the only component that contributes a scalar is the mixed index
\begin{equation}
    R_{\nu j}{}^{\mu i}(\Omega_+)=-\ft12\nabla_\nu N_{-}^{\mu i}{}_j.
\label{eq:linriemann}
\end{equation}
Moreover, the $t_{8} t_{8} R(\Omega_+)^{4}$ term in \eqref{t8 t8 expanded} contains one $t_8$ contracted with the $i$-type indices while the other is contracted with the $j$-type indices. Then, the resulting four-point expression automatically takes the form of a trace over $\nabla N_\pm$ with alternating signs, written schematically as
\begin{align}
    t_{8} t_{8} R(\Omega_+)^{4} \sim \Tr(\nabla N_{+}\nabla N_{-}\nabla N_{+}\nabla N_{-})+\Tr(\nabla N_{+}\nabla N_{-})\Tr(\nabla N_{+}\nabla N_{-}) + \cdots,
\end{align}
where we have omitted coefficients as well as the spacetime indices, and the ellipses indicate terms beyond the four-point function.

On the other hand, from the structure of $\epsilon_8\epsilon_8 R(\Omega_+)^4$ in \eqref{eq:e8e8R4x}, we see that some of the $i$-type indices in (\ref{eq:linriemann}) will be contracted with $j$-type indices.  This would lead to combinations of $\nabla N_\pm$ where the signs do not alternate, hence giving an expression that is not manifestly T-duality invariant
\begin{equation}
    \epsilon_8\epsilon_8 R(\Omega_+)^4 \sim \Tr(\nabla N_{1}\nabla N_{2}\nabla N_{3}\nabla N_{4}) + \cdots.
\label{eq:noTdual?}
\end{equation}
Here the subscripts denote the $\pm$ sign of $N$, and in some cases they do not alternate, signifying a potential issue with T-duality.  The resolution of this T-duality puzzle is that the $\epsilon_8\epsilon_8 R(\Omega_+)^4$ term is actually a five-point (and higher) coupling.  Focusing only on the linearized mixed component Riemann, (\ref{eq:linriemann}), we can write $\epsilon_8\epsilon_8R^4$ in an unexpanded form
\begin{align}
    \epsilon_8\epsilon_8R(\Omega_+)^4&=\fft1{16}\epsilon_{\mu_1\mu_2\mu_3\mu_4}\epsilon^{\nu_1\nu_2\nu_3\nu_4}\epsilon_{i_1i_2i_3i_4}\epsilon^{j_1j_2j_3j_4}\nn\\
    &\kern4em\times(\nabla_{\nu_1}N_-^{\mu_1i_1}{}_{j_1})(\nabla_{\nu_2}N_-^{\mu_2i_2}{}_{j_2})(\nabla_{\nu_3}N_-^{\mu_3i_3}{}_{j_3})(\nabla_{\nu_4}N_-^{\mu_4i_4}{}_{j_4}).
\end{align}
Making use of the antisymmetry of $\epsilon_{\mu\nu\rho\sigma}$, we can rewrite this as
\begin{align}
    \epsilon_8\epsilon_8R(\Omega_+)^4&=\fft1{16}\epsilon_{\mu_1\mu_2\mu_3\mu_4}\epsilon^{\nu_1\nu_2\nu_3\nu_4}\epsilon_{i_1i_2i_3i_4}\epsilon^{j_1j_2j_3j_4}\nn\\    &\kern4em\times[\nabla_{\nu_1}(N_-^{\mu_1i_1}{}_{j_1}(\nabla_{\nu_2}N_-^{\mu_2i_2}{}_{j_2})(\nabla_{\nu_3}N_-^{\mu_3i_3}{}_{j_3})(\nabla_{\nu_4}N_-^{\mu_4i_4}{}_{j_4}))+\cdots],
\label{eq:tote8e8}
\end{align}
where the additional terms are five-point couplings of the form $RN^2(\nabla N)^2$ with the Riemann tensor being obtained from the commutator of two covariant derivatives acting on $N$.  When this is inserted into the tree-level effective action, the total derivative term can be integrated by parts.  This hits the dilaton factor, $e^{-\Phi}$, and hence is also a five-point coupling of the form $N(\nabla\Phi)(\nabla N)^3$.  The end result is that $\epsilon_8\epsilon_8 R(\Omega_+)^4$ does not contribute to the four-point contact term, although it will become important at the five-point level and beyond.

Returning to the form of the expanded $\epsilon_8\epsilon_8R(\Omega+)^4$ term in (\ref{eq:e8e8R4x}), it is still the case that a straightforward reduction would yield a sum of terms of the form (\ref{eq:noTdual?}).  However, as in (\ref{eq:tote8e8}), a judicial rearrangement of terms would allow it to be rewritten as a total derivative plus higher-point contributions
\begin{equation}
    \epsilon_8\epsilon_8 R(\Omega_+)^4 \sim \nabla\Tr( N_{1}\nabla N_{2}\nabla N_{3}\nabla N_{4}) + \cdots.
\end{equation}
This indicates that the generic procedure outlined in Section~\ref{subsection: General approach to verifying T-duality}, starting with the four-point function and rewriting $(\nabla N)^4\to(\nabla\nabla\mathcal S)^4$, must be applied judiciously, as in some cases $(\nabla N)^4$ will instead be pushed to a higher-point coupling through field redefinitions.

At the four-point level, we have verified that the $t_8t_8R(\Omega_{+})^4$ coupling is compatible with T-duality, while no constraints are placed on $\epsilon_8\epsilon_8R(\Omega_{+})^4$.  Alghough perhaps straightforward, this is nevertheless a non-trivial check since the four-point function is non-vanishing for the torus reduction.  This is in contrast to the cosmological reduction where the only possible derivative is the time derivative.  In that case, $\dot N$ can be replaced schematically by $N^2$ using the lower-order equations of motion.  Thus the four-point function vanishes, and non-trivial checks can only be performed at the level of the eight-point function of order $N^8$.


\section{Discussion} \label{section: discussion}

We have investigated T-duality building blocks in the context of higher derivative corrections. With a torsionful connection, the linear combination of the metric and B-field readily appear in $O(d,d)$ invariants. Indeed, this has already been shown at order $\alpha'$ and $\alpha'^3$ in the cosmological reduction \cite{David:2021jqn} and this work generalizes the framework for $d$-dimensional compactifications. The use of the matrix $N_\pm$ makes manifest which terms can be T-duality invariant from the order of the signs for each product of traces that appear in the action. This also suggests that there are in fact some hidden generalized geometrical structures that appear when there is a nonzero torsion.

We have then used these building blocks to construct the first order $\alpha'$ corrections to both the heterotic and bosonic string action. Field redefinitions were required to write the action in a canonical form of the form $\Tr(N^4)$. We find that at this order the T-duality invariant action is written in terms of the trace of the products of $N$ matrices with alternating signs. However, in contrast to the cosmological reduction the two actions do not have the same form as the linear combination of the $O(d,d)$ invariants are not the same.

There are several interesting open problems in the context of constructing these T-duality building blocks when the torsion is nonzero.  Our general framework has been to start with a set of higher-derivative couplings and then to reduce them on a torus, yielding expressions written in terms of traces of $N_\pm$ and $\nabla N_\pm$.  We then aim to rewrite these expressions using the manifestly $O(d,d)$ scalar matrix $\mathcal S$ and its derivatives.  However, ideally, we would like to start with manifestly T-dual building blocks made out of traces of $\nabla\mathcal S$ and $\nabla\nabla\mathcal S$ and then lift them to a corresponding set of ten-dimensional higher-derivative couplings.  Of course, it is easy to go from $\mathcal S$ to $N_\pm$.  But going from $N_\pm$ to covariant and gauge invariant ten-dimensional expressions looks to be highly non-trivial.

We have also focused only on the reduced scalar sector, while ignoring the Kaluza-Klein and winding gauge fields. T-duality acting on the gauge fields can be highly constraining. Reduction on a single circle is sufficient to constrain the full set of NSNS couplings of the type II string at $\mathcal O(\alpha'^3)$, provided one makes full use of the gauge fields \cite{Garousi:2020gio}.  The $\mathcal O(\alpha')$ reduction with gauge fields was worked out in \cite{Eloy:2020dko} for the bosonic and heterotic cases.

One way to make T-duality manifest is through DFT, which involves manifest $O(D,D)$ invariant objects with $D=10$ or $D=26$.  While we have not taken a DFT approach, and instead chose to focus on the decomposition into $N_\pm$ matrices, it is worth emphasizing that DFT leads to an elegant description of the $\mathcal O(\alpha')$ \cite{Marques:2015vua,Baron:2017dvb} and $\mathcal O(\alpha'^2)$ \cite{Hronek:2021nqk,Hronek:2022dyr} invariants.   However, an obstruction arises at $\mathcal O(\alpha'^3)$ \cite{Hronek:2020xxi}.  Although an $O(d,d)$ invariant certainly exists at this order, it cannot be made $O(D,D)$ invariant in the language of DFT.  It would be curious to see if this has any implications in connecting the torus reduced expressions involving $N_\pm$ with the T-dual invarints built out of traces of $\mathcal S$ and its derivatives when going beyond $\mathcal O(\alpha'^2)$.

Finally, although the four-point couplings in the type II case are easily explored, we would naturally wish to extend the analysis to higher-point couplings.  This was done for the cosmological reduction case all the way to the eight-point coupling in the absence of the B-field \cite{Codina:2020kvj} and up to order $H^2R^3$ when the B-field is included \cite{David:2021jqn}.  While the expressions for a $d$-dimensional reduction are more challenging to manipulate, the five-point couplings not involving the dilaton were determined in \cite{Wulff:2021fhr}, and perhaps the higher-point couplings could also be tackled.  This would already be sufficient to yield new insights on the structure of the $\mathcal O(\alpha'^3)$ couplings of the type II string.

\section*{Acknowledgements}

We wish to thank N. Bobev and L. Wulff for enlightening discussions.  This work was supported in part by the U.S. Department of Energy under grant DE-SC0007859. M.D. is supported by KU Leuven C1 grant ZKD1118 C16/16/005, and by the Research Programme of The Research Foundation – Flanders (FWO) grant G0F9516N.

\bibliographystyle{JHEP}
\bibliography{references}
\end{document}